\documentstyle[epsfig]{aipproc}

\begin{document}
\title{Beaming and Jets in Gamma Ray Bursts}
 \author{Re'em Sari}
\address{Theoretical Astrophysics 130-33, California institute of Technology, Pasadena CA 91125}
\maketitle

\begin{abstract}
The origin of GRBs have been a mystery for almost 30 years. The
afterglow observed in the last few years enabled redshift
determination for a handful of bursts, and the cosmological origin is
now firmly established. Though the distance scale is settled, there
still remains orders of magnitude uncertainty in their rate and in the
total energy that is released in the explosion due to the possibility
that the emission is not spherical but jet-like. Contrary to the GRB
itself, the afterglow can be measured up to months and even years
after the burst, and it can provide crucial information on the
geometry of the ejecta. We review the theory of afterglow from jets
and discuss the evidence that at least some of the bursts are not
spherical. We discuss the prospects of polarization measurements, and
show that this is a powerful tool in constraining the geometry of the
explosion.
\end{abstract}

\section{Jets? - A fundamental question}

The study of $\gamma$-ray bursts was revolutionized when the
Italian-Dutch satellite BeppoSAX delivered arcminutes positioning of
some GRBs, within a few hours after the event. This enabled other ground
and space instruments to monitor the relatively narrow error
boxes. Emission in X-ray, infrared, optical and radio, so called
``afterglow'', was observed by now for more than a dozen of bursts.

The current understanding of the GRBs phenomenon is that a compact source
emits relativistic flow with Lorentz factor $\gamma$ of at least a few
hundreds. This flow emits, probably by internal shocks (see e.g. \cite{SP97,FMN96}),
the GRB. After these internal shocks have produced the GRB, the ultra
relativistic flow interacts with the surrounding medium and
decelerates. Synchrotron radiation is emitted by the heated
surrounding matter. As more and more of the surrounding mass is
accumulating, the flow decelerates and the emission shifts to lower
and lower frequencies. Excitingly, the afterglow theory is relatively
simple. It deals with the emission on timescales much longer than those
of the GRBs. The details of the complex initial conditions are
therefore forgotten and the evolution depends only on a small number of
parameters.

We begin by clarifying some of the confusing terminology. There are
two distinct, but related, effects.  The first, {\bf ``jets''},
describes scenarios in which the relativistic flow emitted from the
source is not isotropic but collimated towards a finite solid
angle. The term jet refers to the geometrical shape of the
relativistic flow emitted from the inner engine. The second effect is
that of {\bf ``relativistic beaming''}. The radiation from any object
that is radiating isotropically in its own rest frame, but moving with
a large Lorentz factor $\gamma$ in the observer frame, is collimated
into a small angle $1/\gamma$ around its direction of motion. This is
an effect of special relativity. It has nothing to do with the
ejecta's geometry (spherical or jet) but only with the fact that the
ejecta is moving relativisticly.  The effect of relativistic beaming
allows an observer to see only a small angular extent, of size
$1/\gamma$ centered around the line of sight. Unfortunately, the term
beaming was also used for ``jets'' by many authors (including myself).
We will keep a clear distinction between the two in this paper.  Since
we know the flow is ultra-relativistic (initially $\gamma>100$), there
is no question that the relativistic beaming is always relevant for
GRBs. The question we are interested in is that of the existence of
``jets''.

\begin{figure*}
\centerline{\epsfig{file=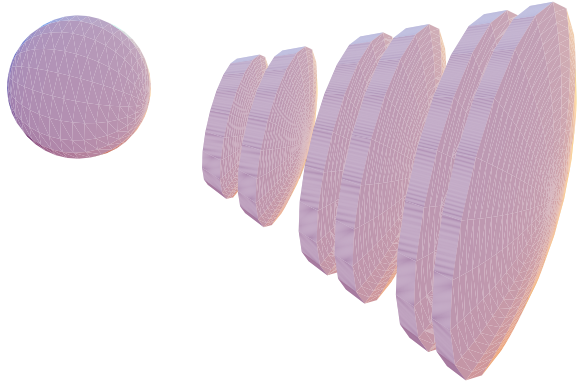,width=3.0in}}
\vspace{10pt}
\caption{Schematic geometric description of jets in GRBs. The scheme shows
the multiple shells before internal shocks have occurred. After that they 
all merge to one shells with typical width a factor of $\gamma^2$ thinner
than their distance from the source. 
}
\label{grbjets}
\end{figure*}

The idealized description of a jet is a flow that occupies only a
conical volume with half opening angle $\theta_0$. In fact the
relativistic dynamics is such that the width of the matter in the
direction of its propagation is much smaller than its distance from
the source by a factor of $1/\gamma^2$. The flow, therefore, does not
fill the whole cone. Instead it occupies only a thin disk at its base,
looking more like a flying pancake \cite{P99} - see figure
\ref{grbjets}.  If the ``inner engine'' emits two such jets in
opposite directions then the total solid angle towards which the flow
is emitted is $\Omega=2\pi \theta_0^2$. The question whether the
relativistic flow is in the form of a jet or a sphere has three
important implications.
 
\noindent {\bf The Total Emitted Energy.} Optical observations of
afterglows enabled redshift determination, and therefore a reasonably
accurate estimate of the distance, $D$, to these events (the
uncertainty is now in the cosmological parameters of the
universe). The so called ``isotropic energy'' can then be inferred
from the fluence $F$ (the total observed energy per unit area at
earth) as $E_{iso}=4\pi D^2 F$ (taking cosmological corrections into
account, $D=D_L/\sqrt{1+z}$ where $D_L$ is the luminosity distance and
$z$ is the redshift). The numbers obtained in this way range from
$10^{51}$erg to $10^{54}$erg with the record of $3\times 10^{54}$erg
held by the famous GRB~990123. These huge numbers approach the
equivalent energy of a solar mass, all emitted in a few tens of
seconds!

These calculations assumed that the source emitted the
same amount of energy towards all directions. If instead the emission
is confined to some solid angle $\Omega$ then the true energy is
$E=\Omega D^2 F$. As we show later $\Omega$ is very weakly constrained
by the GRB itself and can be as low as $10^{-6}$. If so the true
energy in each burst $E \ll E_{iso}$. We will show later that
interpretation of the multi-wavelength afterglow lightcurves indeed indicates
that some bursts are jets with solid angles considerably less than $4 \pi$.
The isotropic energy estimates may be fooling us by a few orders of 
magnitudes! Clearly this is of fundamental importance when considering 
models for the sources of GRBs.

\noindent {\bf The Event Rate.} BATSE sees about one burst per
day. With a few redshifts measured this translates to about $10^{-7}$
bursts per year per galaxy. However,
if the emission is collimated to $\Omega \ll 4\pi$ then we do not see
most of the events. The true event rate is then larger than that measured by
BATSE by a factor of $4\pi /\Omega$. Again this is of fundamental importance. 
Clearly, the corrected GRB
event rate must not exceeds that of compact binary mergers or 
the birth rate of massive stars if these are to produce the majority 
of the observed GRBs.

\noindent {\bf The Physical Ejection Mechanism.} Clearly, different
physical models are needed to explain collimated and isotropic
emission.  For example, in the collapsar model (e.g. \cite{MW99}), 
relativistic ejecta
that is capable of producing a GRB is produced only around the
rotation axis of the collapsing star with half opening angle of about
$\theta_0 \cong 0.1$. Such models would have difficulties to explain
isotropic bursts as well as very narrow jets.

With these uncertainties we are therefore left with
huge ignorance in how, how much and how many GRBs are produces. The
question as to whether the emission of GRBs is spherical or collimated in jets is fundamental to almost all aspects of the GRB
phenomenon.

\section{Afterglow Spectrum - Basic Theory}

\begin{figure*}
\centerline{\epsfig{file=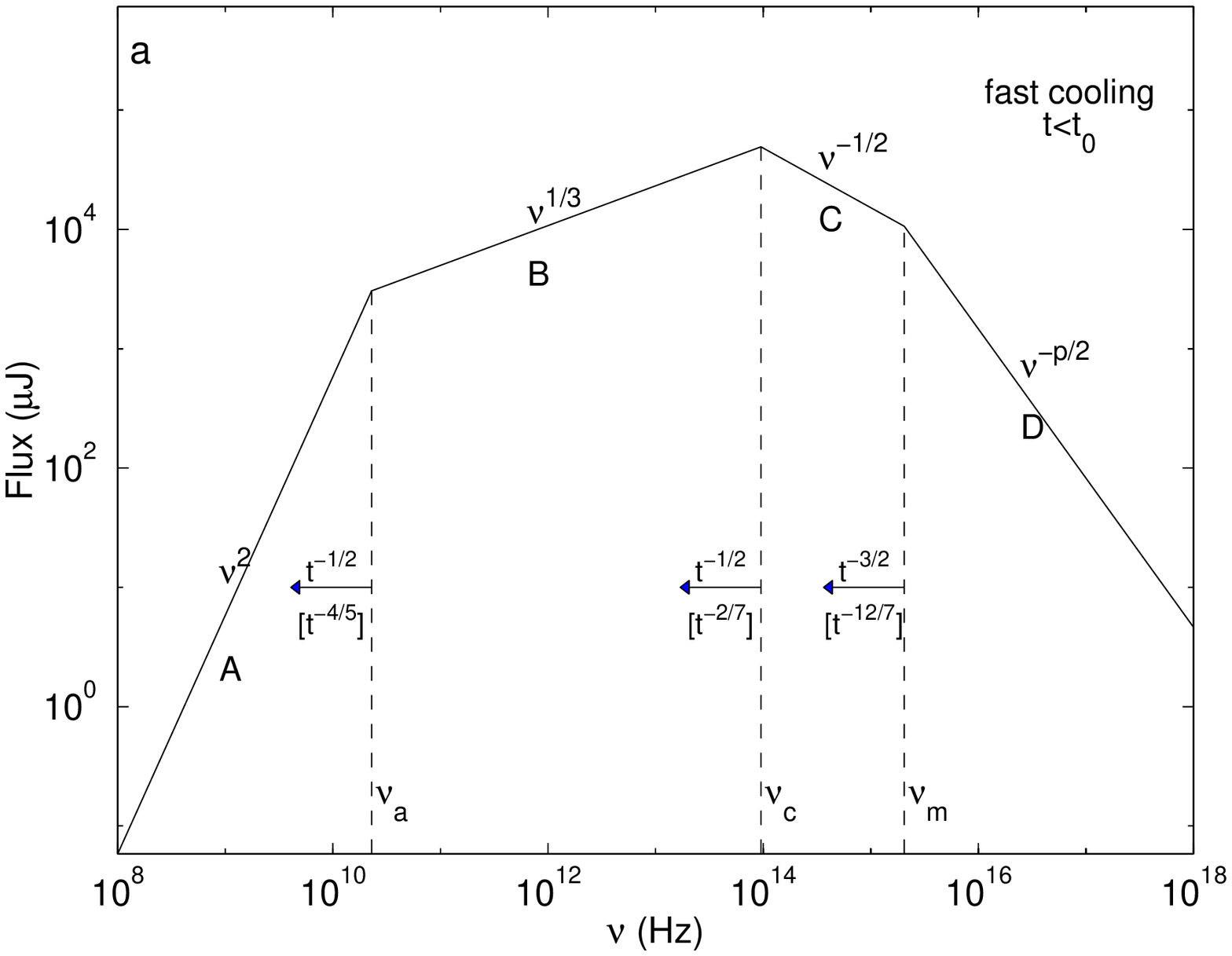,width=2.8in}\ \
\epsfig{file=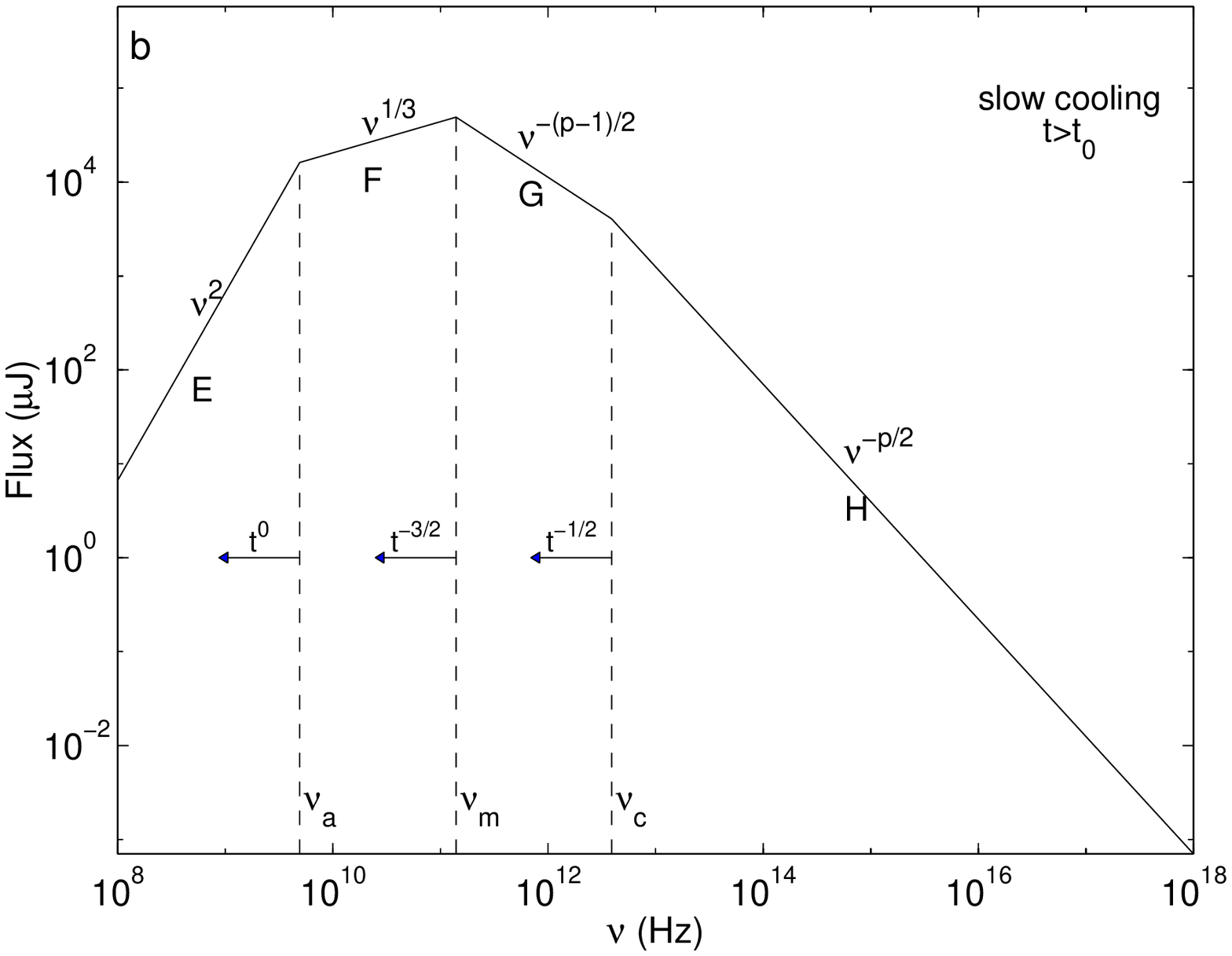,width=2.8in}}
\vspace{10pt}
\caption{Theoretical spectra of synchrotron emission from fast cooling
($\nu_c<\nu_m$ left) and slow cooling ($\nu_m<\nu_c$ right) powerlaw
distribution of electrons. This spectrum is robust and holds for jets
as well as spherical ejecta. In general, the break frequencies change
in time as well as the overall normalization. The arrows on the figure
indicate the evolution of these break frequencies for a spherical
emission in a constant density environment. $p=2.2-2.4$ fits the
observed spectra well.}
\label{SPNspec}
\end{figure*}

When the ejecta interacts with the surrounding medium, a shock waves
(so called the forward shock) is going through the cold ambient medium
and heating it up to relativistic temperatures.  The basic afterglow
model assumes that electrons are accelerated by the shock into a
powerlaw distribution of their Lorentz factor $\gamma_e$: $N(\gamma
_{e})\sim \gamma _{e}^{-p}$ for $\gamma _{e}>\gamma _{m}$. The lower
cutoff of this distribution is assumed to be a fixed fraction of
equipartition. It is also assumed that a considerable magnetic field
is being built behind the shock, again characterized by a certain
fraction of equipartition. The relativistic electrons then emit
synchrotron radiation and produce the observed afterglow. The broad
band spectrum of such emission was given by Sari, Piran \& Narayan
\cite{SPN98} (see figure \ref{SPNspec}).

At each instant, there are three characteristic frequencies: (I) $\nu
_{m}$ which is the synchrotron frequency of the minimal energy
electron, having a Lorentz factor $\gamma _{m}$. (II) The cooling time
of an electron is inverse proportional to its Lorentz factor $\gamma
_{e}$.  Therefore, electrons with a Lorentz factor higher than some
critical value $\gamma _{e}>\gamma _{c}$ can cool on the dynamical
timescale of the system. This characteristic Lorentz factor
corresponds to the ``cooling frequency'' $\nu _{c}$. (III) Below some
critical frequency $\nu _{a}$ the flux is self absorbed and is given
by the Rayleigh-Jeans portion of a black body spectrum. The broad band
spectrum of the well studied GRB 970508 \cite{G+98} is in very good
agreement with the theoretical picture.

We stress that the spectrum given above is quite robust. The only
assumption is synchrotron radiation from a powerlaw distribution of
relativistic electrons. The same spectrum will hold whether the shocks
propagates into a constant density interstellar medium or a decreasing
surrounding density produced earlier by the progenitor's wind. It will
be valid whether the ejecta is spherical or jet-like, whether the
equipartition parameters are constant with time or not.

On the contrary, the temporal evolution of the spectrum is more subtle. 
The simplest evolution, which well describes the
data of some bursts, is the spherical adiabatic model with a constant
density ambient medium. In this scenario, $\gamma \sim R^{-3/2}$
or in terms of the observer time, $t=R/\gamma^2c$,  $\gamma \sim
t^{-3/8}$. Given the evolution of $\gamma(t)$ one can derive the 
temporal evolution of the break frequencies and the results are indicated
in figure \ref{SPNspec}. The peak flux, in the adiabatic, spherical 
constant ambient density model is constant with time.

\section{Hydrodynamics of Jets}

Interestingly, due to the effect of relativistic beaming (which is independent
of jets) we are only able to see an angular extent of $1/\gamma <0.01$
during the GRB itself where the Lorentz factor $\gamma
>100$. Moreover, it is only regions of size $1/\gamma$ that are
causally connected. Therefore, each fluid element evolves as if it is
part of a sphere as long as $1/\gamma<\theta_0$. Combining these two
facts, we cannot distinguish a jet from spherical ejecta as long as
$1/\gamma<\theta_0$.

However, as the afterglow evolves, $\gamma$ decreases and it will eventually
fall below the initial inverse opening angle of the jet. The observer
will notice that some of the sphere is missing from the fact that
less radiation is observed. This effect alone, will produce a significant
break, steepening the lightcurve decay by a factor of $\gamma^2 \sim
t^{-3/4}$ even if the dynamics of each fluid element has not
changed. The transition should occur at the time $t_{jet}$ when
$1/\gamma \cong \theta_0$. Observing this time can therefore provide
an estimate of the jet's opening angle according to
\begin{equation}  \label{t_jet}
t_{{\rm jet}}\approx 6.2 (E_{52}/n_{1})^{1/3}(\theta _{0}/0.1)^{8/3}
{\rm hr}.
\end{equation}

Additionally, Rhoads \cite{R99} has shown that at about the same time
(see however \cite{PM99,MR99,MSB99}),
the jet will begin to spread laterally so that its opening angle 
$\theta (t\grave{)}\sim 1/\gamma$. 
The ejecta now encounters more surrounding matter and
decelerates faster than in the spherical case. The Lorentz factor 
now decays exponentially with the radius and as $\gamma\sim t^{-1/2}$ 
with observed time. Taking this into account, the observed 
break is even more significant. The slow cooling spectrum given in figure 
\ref{SPNspec} evolves now with decreasing peak flux $F_{\nu,m} \sim t^{-1}$ 
and the break frequencies evolve as $\nu_m \sim t^{-2}$, $\nu_c \sim t^0$ and
$\nu_a \sim t^{-1/5}$. 
This translate to a temporal decay in a given frequency as
listed in table \ref{t:afterglow}.

\begin{table}[b]
\begin{tabular}{|c||c||c|c|}
\hline
& spectral index & \multicolumn{2}{|c|}{light curve index $\alpha$, $F_{\nu}\propto
t^{-\alpha}$} \\
& $\beta$, $F_{\nu}\propto \nu^{-\beta}$ & sphere & jet \\ \hline\hline
$\nu<\nu_a$ & $\beta=-2$ & $\alpha=-1/2$ & $\alpha=0$ \\ \hline
$\nu_a<\nu<\nu_m$ & $\beta=-1/3$ & $\alpha=-1/2$ & $\alpha=1/3$ \\ \hline
&  & $\alpha=3(p-1)/4\cong 1.05$ & $\alpha=p\cong 2.4$ \\
\raisebox{1.5ex}[0pt]{$\nu_m<\nu<\nu_c$} & \raisebox{1.5ex}[0pt]{$(p-1)/2
\cong0.7$} & $\alpha=3\beta/2$ & $\alpha=2\beta+1$ \\ \hline
&  & $\alpha=(3p-2)/4\cong 1.3$ & $\alpha=p\cong 2.4$ \\
\raisebox{1.5ex}[0pt]{$\nu>\nu_c$} & \raisebox{1.5ex}[0pt]{$ p/2
\cong1.2$} & $\alpha=3\beta/2-1/2$ & $\alpha=2\beta$ \\ \hline
\end{tabular}
\caption{The spectral index $\beta$ and the temporal index $\alpha$ as
function of $p$ for a spherical and a jet-like evolution. Typical values are 
quoted using $p=2.4$. The parameter free relation between $\alpha$ and 
$\beta$ is given for each case (eliminating $p$). The difference in $\alpha$
between a jet and a sphere is always substantial at all frequencies.}
\label{t:afterglow}
\end{table}

The jet break is a hydrodynamic one. It should therefore appear at
the same time at all frequencies - an achromatic break. Though an achromatic
break is considered to be a strong signature of a jet, one should keep
in mind that any other hydrodynamic transition will also produce an
achromatic break. To name a few: the transition from relativistic to 
non-relativistic dynamics, a jump in the ambient density or 
the supply of new energy
from slower shells that catch up with the decelerated flow. However,
the breaks produced by the transition from a spherical like evolution
(when $1/\gamma<\theta_0$) to a spreading jet has a well defined
prediction for the change in the temporal decay indices. The amount of 
break depends on the spectral regime that is observed. It can be seen
from table \ref{t:afterglow} that the break is substantial $\Delta
\alpha >0.5$ in all regimes and should be easily identified.

Finally we note that if jet's opening angle is of order unity, the
total energy may still be about an order of magnitude lower than the
isortropic estimate. However, in this case the break will be
``hidden'' as it will overlap the transition to non-relativistic
dynamics. It was suggested that this is the case for GRB~970508
\cite{FWK99}

\section{Observational Evidence for Jets}

Evidence of a break from a shallow to a steep power law was first seen
in GRB 990123 \cite{K+99a,F+99}. Unfortunately the break was observed
only in one optical band while the infrared data was ambiguous.  Yet,
the strongest evidence for this burst being a jet does not come from
this optical break but rather from radio observations, as explained
below. A famous and exciting event this year was the first detection
of a bright (9th magnitude) optical emission simultaneous with GRB
990123 \cite{A99}. Another new ingredient in GRB 990123 is a radio
flare \cite{K+99b}. Contrary to previous afterglows, where the radio
peaks around few weeks and then decays slowly, this burst had a fast
rising flare, peaking around a day and then decaying quickly. Sari and
Piran \cite{SP99c} have shown that the bright optical flash and the
radio flare are related.  Within a day the emission from the
adiabatically cooling ejecta, that produced the $60$s optical flash
shifts into the radio frequencies. Given this interpretation, the
regular forward shock emission should have come later, on the usuall
few weeks timescale. The fact that this ``usual'' forward shock radio
emission did not show up is in agreement with the interpretation of
this burst as a ``jet'' which causes the emission to considerably
weaken by the time the typical frequency $\nu_m$ arrives to radio
frequencies.
\begin{figure}
\centerline{\epsfig{file=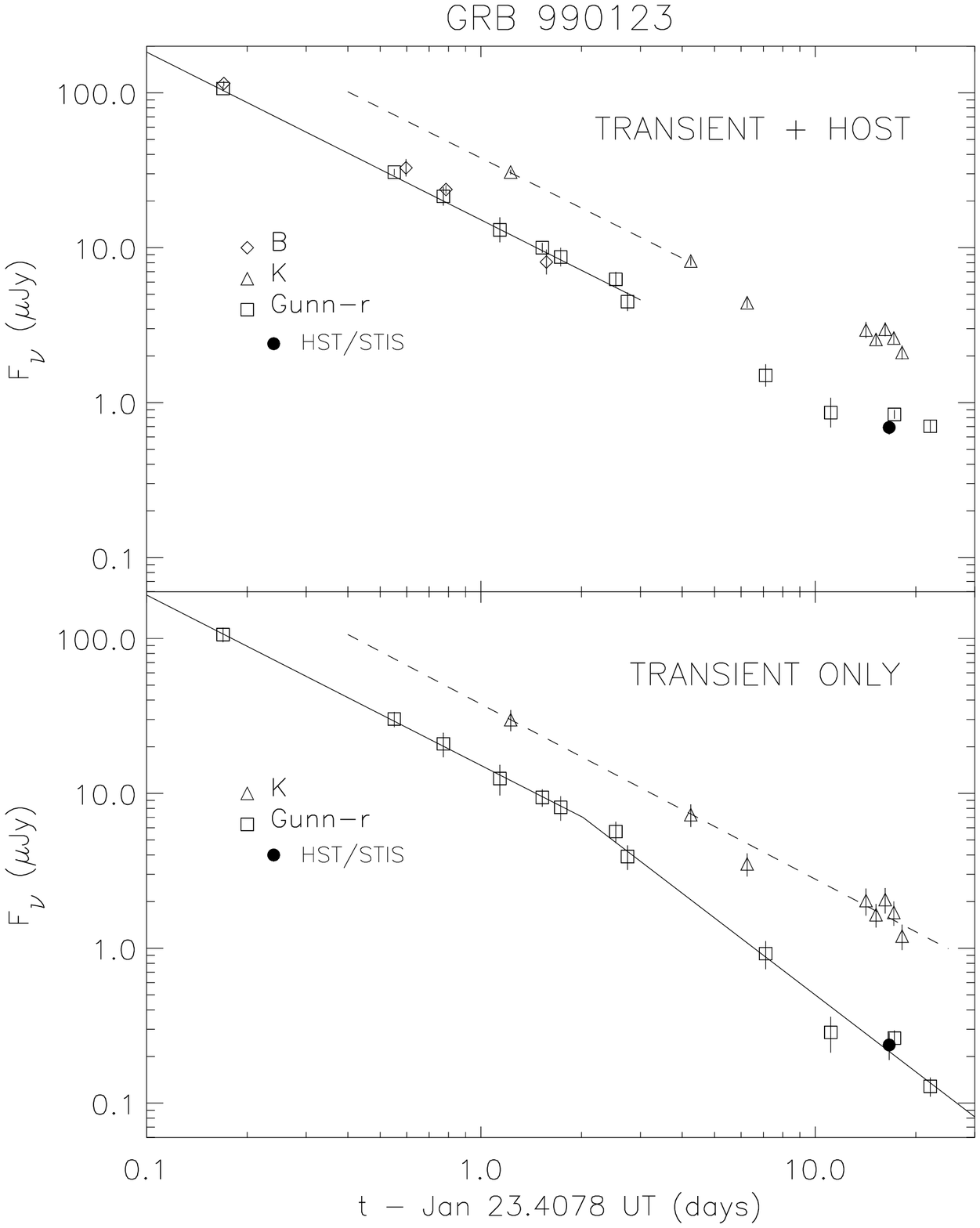,width=2.2in}\ \
\epsfig{file=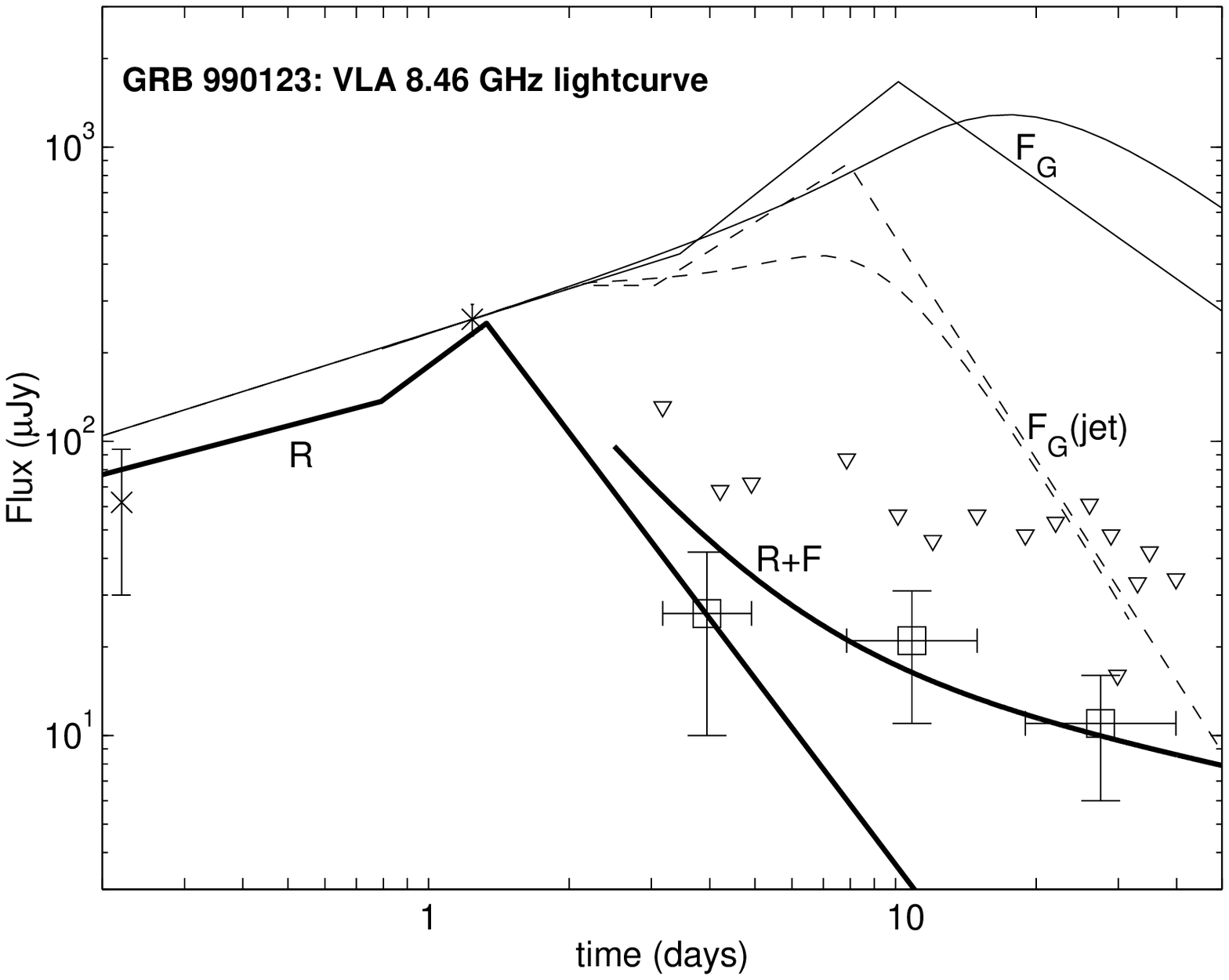,width=3.4in}}
\vspace{10pt}
\caption{GRB~990123: Optical data (left) shows some break in the light
curve at Gunn-r band. K band seems to have no break but the
contribution of the host galaxy is less certain.
Radio ``flare'' (right) seen a day after the burst agrees with 
theoretical scaling 
of the optical flash (heavy solid line marked R).  
In the jet interpretation, only
faint radio emission is expected on late times as given by the heavy
solid line marked R+F, in agreement with observations.  Thin and
dashed lines indicate the theoretical expectations if the radio signal
at day two is interpreted as the forward shock (independent of the optical flash) and if jets are not
taken into account. These will largely over predict the late radio 
upper limits, marked by triangles \protect\cite{K+99b} (see however 
\protect\cite{G+99}).}
\end{figure}

\begin{figure}
\centerline{\epsfig{file=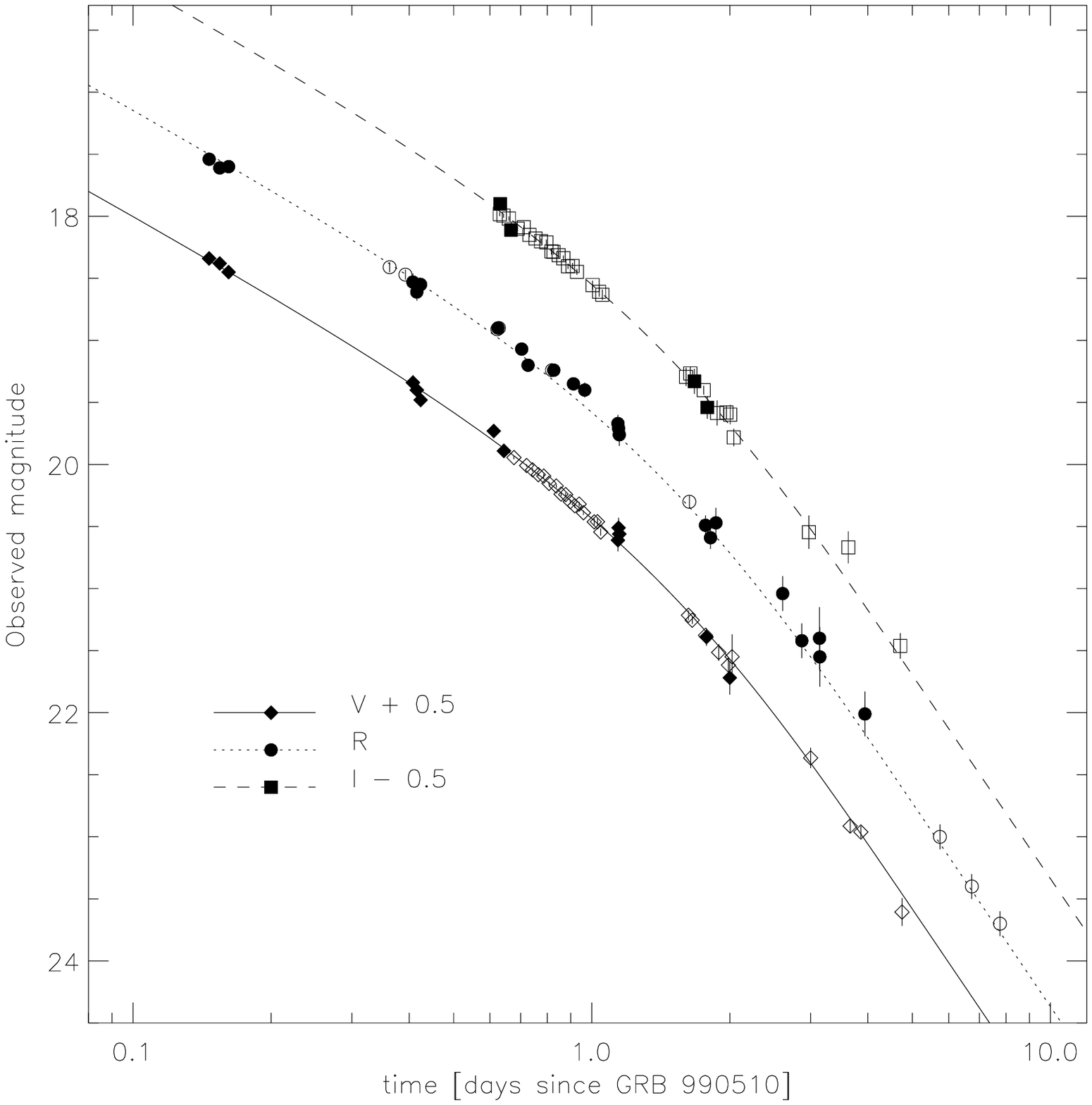,width=2.6in}\ \ 
\epsfig{file=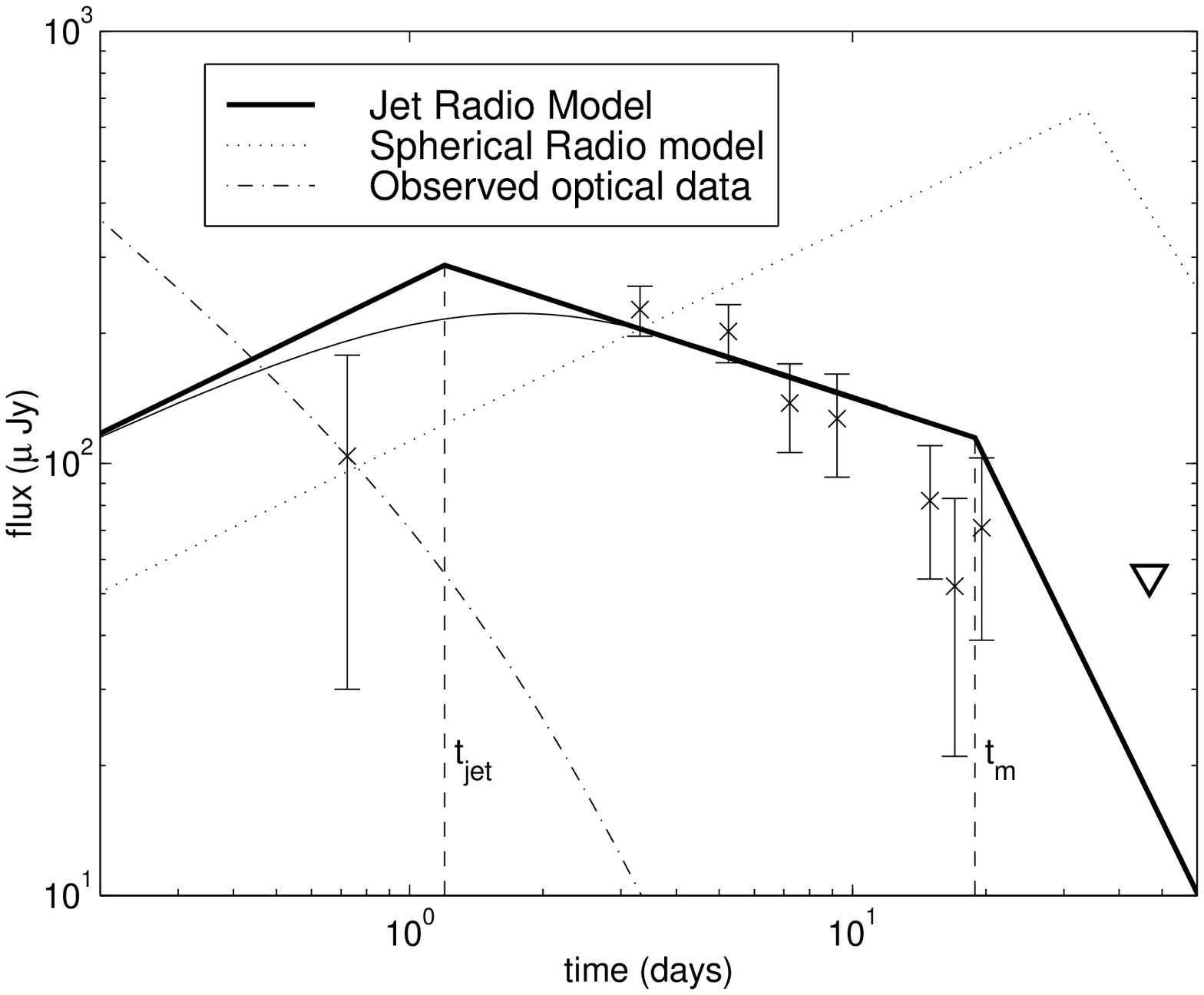,width=3.0in}}
\vspace{10pt}
\caption{GRB~990510, the best evidence for a ``jet'': an achromatic
break in optical and radio at $t_{jet} = 1.2$ days implying
$\theta_0 = 0.08$. The temporal slope before and after the break
agree well with theory if $p=2.2$. For this burst
$E_{iso}=2.9\times 10^{53}$erg but the true total energy is only
$E=10^{51}$erg.
}
\end{figure}

GRB~990510 had a very clear break simultaneously in all optical bands
and in radio \cite{S+99,H+99}. In GRB~990123 and GRB~990510 the
transition times were about $2.1$ days and $1.2$ days reducing the
isotropic energy estimate by a factor of $\sim 200$ and $\sim 300$,
respectively. The total energy is now well below a solar rest mass!

Sari, Piran \& Halpern \cite{SPH99} have noted that the observed
decays in GRB afterglows that do not show a break are either of a
shallow slope of $\sim t^{-1.2}$ or a very steep slope of $\sim
t^{-2.1}$. They argued that the rapidly decaying bursts are those in
which the ejecta was a narrow jet and the break in the light curve was
before the earliest observation.  Interestingly, evidence for jets are
found when the inferred energy $E_{iso}$ (which does not take jets
into account) is the largest. This implies that jets may account for a
considerable fraction of the wide luminosity distribution seen in
GRBs, and that the true energy distribution is less wide than it seems
to be.

An alternative explanation for these afterglows with fast decline
is propagation into a medium with decreasing density, i.e. a wind produced earlier by the progenitor \cite{CL99}. We favor the jet 
interpretation for two reasons: (I) decreasing density only 
enhance the decay by $t^{-1/2}$ for $\nu_m<\nu<\nu_c$ and does 
not enhance the decay at all for $\nu>\nu_c$ 
(with typical parameters the optical 
and certainly the x-ray bands are above $\nu_c$). The rest of the needed 
effect, in the wind interpretation, is associated with a higher value of
the electron powerlaw distribution index $p$ ($p \cong 3$ instead of
$p\cong 2.2-2.4$). Why should the value of $p$ be different for shocks
propagating into winds? With the jet interpretation one can explain all
afterglows with a single value of $p$, as in \cite{SPH99}. (II) The
jets interpretation makes the luminosity distribution of GRBs more narrow,
since evidence for jets are found in bright events.
Clearly, these are circumstantial evidence. A more clear cut between these
two possible interpretations can be done with the use of early afterglow
observation, preferably at radio frequencies (see \cite{FKS+99}).

In summary, there are several kind of afterglows:

\noindent
{\bf Shallow decline} $\sim t^{-1.2}$ for as long as the afterglow
can be observed. These are probably spherical or at least have a large 
opening angle (e.g. GRB~970508).

\noindent
{\bf Fast decline} $\sim t^{-2.1}$ (e.g. GRB~980519 and GRB~980326). 
These are either narrow jets, in which the break was very early or 
they have high values of $p$ and propagate into decreasing density medium.

\noindent
{\bf Breaks}: Initially slow decline that changes into a fast decline. These
are the best candidates for jets (e.g. GRB~990510).

\section{Polarization - A promising tool}

An exciting possibility to further constrain the models and obtain a
more direct proof of the geometrical picture of ``jets'' is to measure
linear polarization. High levels of linear polarization are usually
the smoking gun of synchrotron radiation.  The direction of the
polarization is perpendicular to the magnetic field and can be as high
as $70\%$.  Gruzinov \& Waxman and Medvedev \& Loeb \cite{GW99,ML99}
considered the emission from spherical ejecta which by symmetry should
produce no polarization on the average, except for fluctuations of
order of a few percent. Polarization is more natural if the ejecta is
a ``jet'' and the line of sight from the observer is with in the jet
but does not coincide with its axis. In this case, the spherical
symmetry is broken \cite{G99,GL99,S99}, and the natural polarization
produced by synchrotron radiation should not vanish. For simplicity,
lets assume that the magnetic field behind the shock is directed along
the shock's plane (the results hold more generally, unless the
magnetic field has no preferred direction). The synchrotron
polarization from each part of the shock front, which is perpendicular
to the magnetic field, is therefore directed radially.

\begin{figure*}
\centerline{\epsfig{file=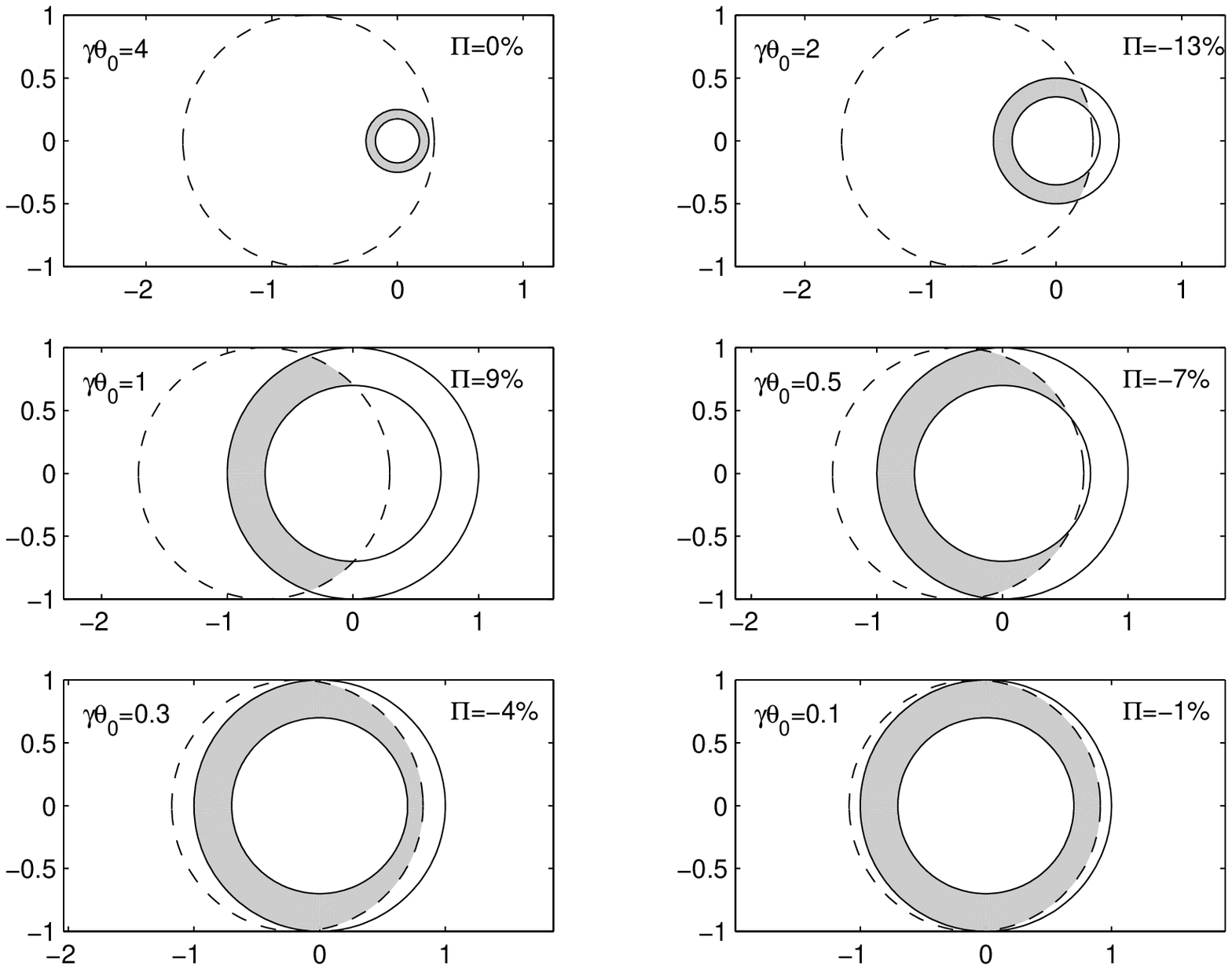,width=2.8in}\ \ \epsfig{file=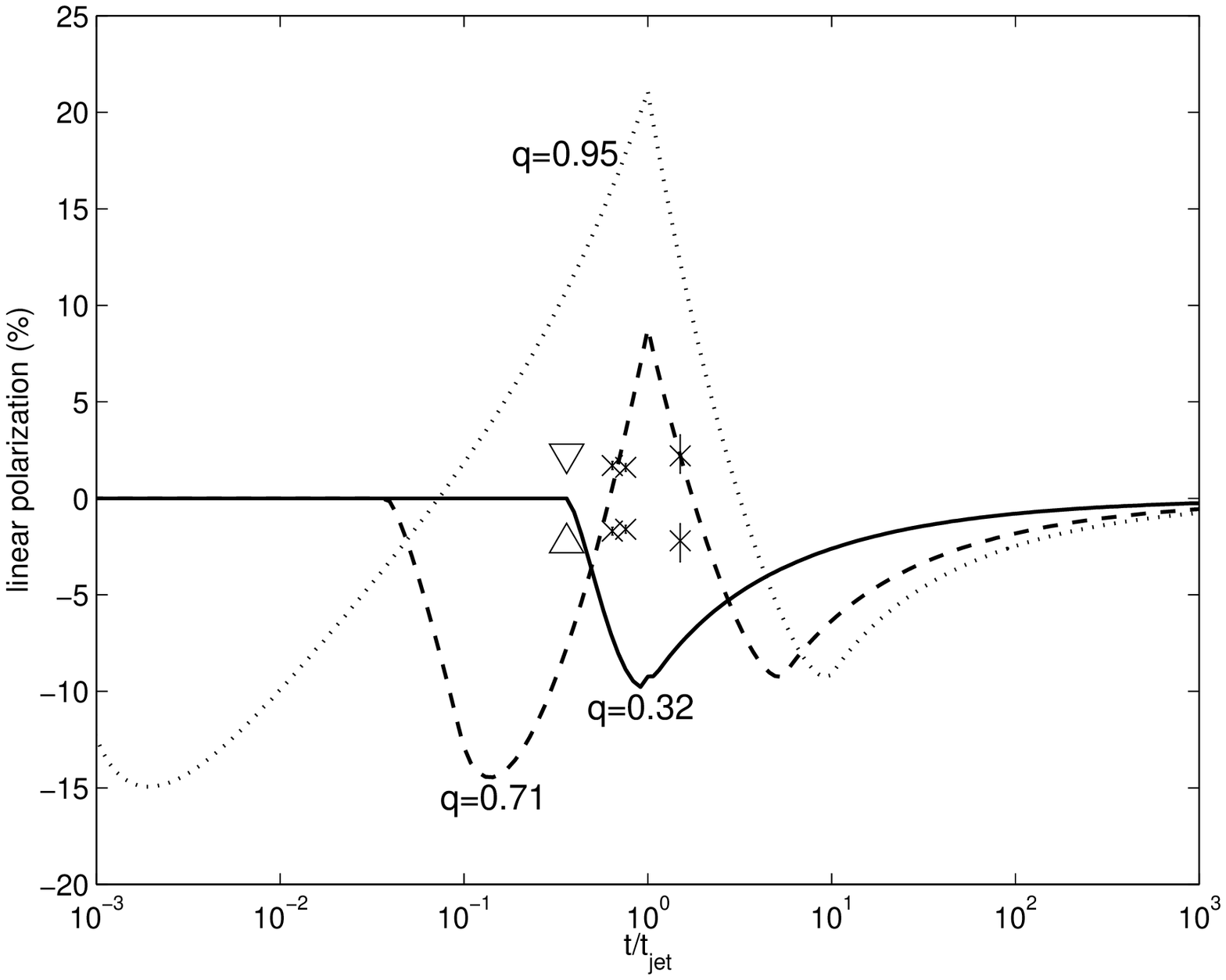,width=2.8in}}
\vspace{10pt}
\caption{Left: Shape of the emitting region.
Dash line marks the physical extent
of the jet while solid lines give the viewable region $1/\gamma$.
The observed radiation is coming from the gray region.
On each frame, the percentage of 
polarization is given on the top right and the initial size of the jet
relative to $1/\gamma$ is given on the left.
The frames are scaled so that the size of the jet is unity.
Right: observed and theoretical polarization lightcurve for three possible
offsets of the observer relative to the jet axis
Observational data for GRB~990510 is marked by x, assuming $t_{jet}=1.2$\,d.
The upper limit for GRB~990123 is given by a triangle, assuming
$t_{jet}=2.1$\,d.}
\label{polfig}
\end{figure*}

As long as the relativistic beaming angle $1/\gamma$ is narrower
than the physical size of the jet $\theta_0$, one is able to see a
full ring and therefore the radial polarization averages to zero 
(the first frame, with $\gamma\theta_0=4$ of the left plot in 
figure \ref{polfig}). As
the flow decelerates, the relativistic beaming $1/\gamma$ becomes
comparable to $\theta_0$ and only a part of the ring is visible; net
polarization is then observed. Note that due to the radial direction
of the polarization from each fluid element, the total polarization is
maximal when a quarter ($\gamma\theta_0=2$ in figure \ref{polfig}) 
or when three quarters ($\gamma\theta_0=1$ in figure \ref{polfig}) 
of the ring are missing (or radiate less efficiently) 
and vanishes for a full and for half ring. The
polarization when more than half of the ring is missing is
perpendicular to the polarization direction when less than half of it
is missing.

At late stages the jet expands and since the offset of the observer
from the physical center of the jet is constant, spherical symmetry is
regained. The vanishing and re-occurrence of significant parts of the
ring results in a unique prediction: there should be three peaks of
polarization, with the polarization position angle during the central
peak rotated by $90^{\circ }$ with respect to the other two peaks. In
case the observer is very close to the center, more than half of
the ring is always observed, and therefore only a single direction of
polarization is expected. A few possible polarization light curve are
presented in figure \ref{polfig}.

\section{Summary}
Now when redshifts for GRBs are routinely measured, the largest
uncertainty in their energy budget and event rate is the possibility
that the emission is not spherical but jet-like. We discussed the
theory of afterglow from jet-like event. These should produce a
substantial break at all frequencies. The time where this break occurs
is an indication of the jets opening angle. GRB~990510 seems to be a
perfect example for this behavior.  The inferred opening angle is
about $0.1$ consistent with upper limits from searches of orphan X-ray
afterglows \cite{GHV+99}. Several other candidate for jets are bursts
with fast decline, where the break presumably took place before the
earliest observation. This question will be settled when more frequent
early observations are available. We have shown that afterglow from
jets should show a unique signature of polarization, at detectable
levels. Observing such signature will confirm the jet interpretation
and the synchrotron model in general.

{\bf Acknowledgements} I thank Titus Galama for very useful comments, and 
the Sherman Fairchild foundation for support.

\end{document}